\documentclass{article}

\usepackage[normalem]{ulem}
\usepackage{lmodern}
\usepackage{subfigure}
\usepackage{caption}

\usepackage[a4paper, total={6in, 8in}]{geometry}

\usepackage{enumitem}
\usepackage{multicol}
\usepackage{graphicx}
\usepackage{epstopdf}
\usepackage{tabularx}
\usepackage[linesnumbered,titlenumbered,ruled,]{algorithm2e}

\usepackage{listings}
\usepackage{lmodern}
\usepackage[usenames,dvipsnames]{xcolor}

\usepackage{algorithmicx}
\usepackage{algpseudocode}
\usepackage{xcolor}

\usepackage{url}

\usepackage[toc,page]{appendix}

\algrenewcommand\alglinenumber[1]{\tiny #1:}
\usepackage{mathtools}
\usepackage{amsmath}%
\usepackage{amsfonts}%
\usepackage{amssymb}%
\usepackage{graphicx}
\usepackage{color}
\usepackage{cases}

\SetAlCapNameFnt{\scriptsize}
\SetAlCapFnt{\scriptsize}
\SetAlgorithmName{Algorithm}{Algo}{List of Algorithms}

\newlist{myenumi}{description}{10}
\setlist[myenumi]{labelindent=\parindent, leftmargin=*, label=(\roman*), align=left}
\setlist[myenumi]{leftmargin=0pt}

\newtheorem{thm}{Theorem}

\newtheorem{proposition}[thm]{Proposition}

\newtheorem{zremark}[thm]{Remark}

\newenvironment{rremark}{\par\small\zremark}{\endzremark}

\newtheorem{innercustomgeneric}{\customgenericname}
\providecommand{\customgenericname}{}
\newcommand{\newcustomtheorem}[2]{%
  \newenvironment{#1}[1]
  {%
   \renewcommand\customgenericname{#2}%
   \renewcommand\theinnercustomgeneric{##1}%
   \innercustomgeneric
  }
  {\endinnercustomgeneric}
}

\newcustomtheorem{conj}{Conjecture}
\newcustomtheorem{customlemma}{Lemma}

\setlist[itemize]{noitemsep, topsep=0pt}

    \newcommand{\iom}{$IO\mathcal{M}$\xspace}
    \newcommand{\fom}{$FO\mathcal{M}$\xspace}
    \newcommand{\foj}{$FO\mathcal{K}_c$\xspace}
    \newcommand{\ioj}{$IO\mathcal{K}_c$\xspace}
    \newcommand{\D}{\Delta }

    \definecolor{darkgreen}{rgb}{0.1, 0.5, 0.1}
\begin{document}

\title{Mandelbrot set and Julia sets of fractional order
}


\author{Marius-F. Danca  \\
STAR-UBB Institute,\\
Babes-Bolyai University,\\
400084, Cluj-Napoca, Romania\\
and\\
Romanian Institute of Science and Technology, \\
400487, Cluj-Napoca, Romania\\
danca@rist.ro\\
and\\
        Michal Fe\v{c}kan \\
             Faculty of Mathematics, Physics and Informatics,\\
              Comenius University in Bratislava, \\84215 Bratislava, Slovakia\\
             and\\
Mathematical Institute, Slovak Academy of Sciences, \\84104 Bratislava, Slovakia\\
michal.feckan@gmail.com
}


\maketitle

\begin{abstract}
In this paper the fractional-order Mandelbrot and Julia sets in the sense of $q$-th Caputo-like discrete fractional differences, for $q\in(0,1)$, are introduced and several properties are analytically and numerically studied. Some intriguing  properties of the fractional models are revealed. Thus, for $q\uparrow1$, contrary to expectations, it is not obtained the known shape of the Mandelbrot of integer order, but for $q\downarrow0$. Also, we conjecture that for $q\downarrow0$, the fractional-order Mandelbrot set is similar to the integer-order Mandelbrot set, while for $q\downarrow0$ and $c=0$, one of the underlying fractional-order Julia sets is similar to the integer-order Mandelbrot set. In support of our conjecture, several extensive numerical experiments were done.
To draw the Mandelbrot and Julia sets of fractional order, the numerical integral of the underlying initial values problem of fractional order is used, while to draw the sets, the escape-time algorithm adapted for the fractional-order case is used. The algorithm is presented as pseudocode.

\vspace{5mm}
\textbf{keywords: }Mandelbrot set of fractional order; Julia set of fractional order; Caputo-like discrete fractional difference
\end{abstract}

\section{Introduction}
The study of the dynamics of complex maps was initiated by P. Fatou and G. Julia in the early of twentieth century, before being revived by B. Mandelbrot.
As know, in the complex plane $\mathbb{C}$, the Integer Order (IO) Mandelbrot set represents the set of complex numbers (parameters) $c$ for which the quadratic map $f_c(z)=z^2 + c$ (Mandelbrot map) does not diverge to infinity when it is iterated with $f_c$ from $z = 0$. This set, has been first defined and drawn by Robert W. Brooks and Peter Matelski in 1978 \cite{mand1}, and later made famous by Benoit Mandelbrot (see e.g. \cite{mand2}). The dynamics generates by $f_c$ represent a huge sources of fractal structures (see e.g. \cite{bibzece,bibunspe,sci,mandelus}).

Julia sets are made of all points which under iterations do not go to an attractor which may be at infinity.
Compared to Mandelbrot set, where $c$ is variable in the parametric plane $\mathbb{C}$, Julia sets are obtained for fixed $c$ and the origin of iterations variable in $\mathbb{C}$.

The Mandelbrot set is known as the set of all points $c$ for which Julia sets are compact and connected.

Details and background on Mandelbrot set and Julia sets can be found in the following works \cite{mand2,dou,bibcinspe,bibsaispe,bibsaptespe,bibzece,devu}, to cite only few of them.

Due the description of memory and hereditary properties, Fractional Order (FO) difference equations still receive increasing attention. However, as mentioned in \cite{88}, overall, fractional calculus, closely related to classical calculus, is not direct generalization of classical calculus in the sense of rigorous mathematics. One of the first definitions of a fractional difference operator has been proposed in 1974 \cite{4}. While there are many works on fractional differential equations, there still are only few works in the theory of the fractional finite difference equations. In \cite{6,mich,micx} problems related to Caputo fractional sums and differences can be found, while in \cite{bibnoua} Initial Value Problems (IVPs) in fractional differences are studied. For stability of fractional differences compare \cite{9,10}, while properties of fractional discrete logistic map, weakly fractional difference equations, symmetry-breaking of fractional maps can be found in \cite{x1,x2,bib1}. The nonexistence of periodic solutions in fractional difference equations is analyzed in \cite{micy}.

Notations utilized in this paper:
\begin{itemize}
\item \iom: Mandelbrot of IO;
\item $IO\mathcal{K}_c$: Filled Julia set of IO;
\item $IO\mathcal{J}_c$: Julia set of IO;
\item \fom: Mandelbrot set of FO;
\item \foj: Filled Julia set of FO;
\item DEM: Distance Estimator Method;
\item IIM: Inverse Iteration Method

\end{itemize}

\section{Mandelbrot set and Julia sets of integer order}

In this section, few of the most important characteristics of $IO\mathcal{M}$ set, and \ioj sets, which will be revealed analytically or numerically in the cases of Mandelbrot and Julia sets of FO, are briefly presented.

The iteration of $f_c$ is obtained by the relation
\begin{equation}\label{zero}
z_{n}=z_{n-1}^2+c, z_0=0, n\in \mathbb{N}^*=\{1,2,...\},
\end{equation}
which generates the sequence
\begin{equation}\label{unu}
z_0=0,z_1=f_c(0)=c,z_2=f_c^2(0)=c^2+c,z_3=f_c^3(0)=(c^2+c)^2+c,...
\end{equation}
where by $f_c^k(0)$ one understands $f_c(f_c^{k-1}(0))$.

A complex number $c\in IO\mathcal{M}$ set if the absolute value of $z_n$ in the sequence \eqref{unu} remains bounded for all $n\in \mathbb{N}^*$, $|z_n|<2$, for all $n\geq0$.

As proposed by Mandelbrot, the boundary of the Mandelbrot set represents a fractal curve (Fig. \ref{fig1} (a)). For the history related to the origin of the \iom set compare \cite{wiki}.

Consider the set of all points $z_0$ which tend to $\infty$ through the iteration \eqref{zero}
\[
A_c(\infty)=\{z_0\in\mathbb{C}: f_c^k(z_0)\rightarrow \infty, \text{~as~} k\rightarrow \infty\}.
\]
The set $A_c(\infty)$ depends on $c$, and his frontier represents the $IO\mathcal{J}_c$ set of $f_c$.

The \emph{filled Julia }set for fixed $c$, of IO, related to $f_c$, $IO\mathcal{K}_c$ which, for computationally reasons is considered in this paper, is the set of all points $z_0\in \mathbb{C}$ for which the orbit \eqref{unu} is bounded
\[
IO\mathcal{K}_c=\mathbb{C}\backslash A_c(\infty)=\{z_0\in \mathbb{C}: f_c^k(z_0) \text{~~remains bounded for all~} k\}.
\]
The $IO\mathcal{J}_c$ set is contained in the $IO\mathcal{K}_c$ set and is the boundary of the $IO\mathcal{K}_c$ set
\[
\partial IO\mathcal{K}_c=IO\mathcal{J}_c=\partial A_c(\infty).
\]

In this paper, beside some new properties related to the Mandelbrot set of FO, the following known properties of Mandelbrot set and Julia sets of IO will be analytically or numerically studied on their FO counterparts.

\begin{enumerate}
\item The \ioj sets for purely real $c$, and \iom set, are symmetric about the real axis (reflection symmetry). Julia sets of $f_c$ are symmetric about the origin.
\item As known, it is conjectured that the Mandelbrot set is locally connected \cite{dou,wiki}, the full conjecture of this very technical and complicated property being still open. Mandelbrot had decided empirically that his isolated islands were actually connected to the mainland by very thin filaments \cite{mil}. In this paper we are interested in the connectivity property as a computationally property which by using, beside the \emph{escape-time} algorithm (see Appendix \ref{ap2}), a performing method (Distance Estimator Method), revel the filaments connecting the apparently isolated islands to the mainland of the Mandelbrot set or Julia sets, of IO or FO. Empirical small areas are considered.

\item The \iom set is the set of all parameters $c$ for which \ioj are connected sets;
\item The \fom set is bounded;
\item For $c$ situated outside of \iom one obtains Cantor sets (``dust'' sets composed of infinitely disjoint points);

\item The Julia sets can be connected, disconnected or totally disconnected (see e.g. \cite{bib2,bib3,bib4}). To note that the \iom set represents the set of all points in the complex plane for which the alternated Julia sets \cite{bib6} are disconnected (but not totally disconnected).
\end{enumerate}

In this paper the filled Julia sets are considered.

Hereafter, by point in the complex plane, one understands the image in the complex plane of a complex number.

Drawing Mandelbrot set and Julia sets of IO and FO, bases on the theorem which states that iterating $f_c$, with starting value $z_0$, only one of the following possibilities happens: either the obtained orbit remains bounded by $2$, or diverges to $\infty$ \cite{sci}.

In Fig. \ref{fig1} (a) is presented the \iom set, while in Figs. \ref{fig1} (c)-(j) there are presented several $IO\mathcal{K}_c$ sets for different $c$ values. In Fig. \ref{fig1} (c), there is presented an $IO\mathcal{J}_c$ set with no interior, with $c=0.359+\i0.599$, while in Fig.\ref{fig1} (d) there are presented comparatively, both the $IO\mathcal{K}_c$ and $IO\mathcal{J}_c$ sets for the same value of parameter $c=0.276+\i0.536$, obtained with the time-escape algorithm and IIM, \cite{sci}, respectively.

\section{Fractional Order Mandelbrot map}\label{sect}

In this section the fractional discretization of the Mandelbrot and Julia sets of FO, in the sense of Caputo-like are introduced.

To obtain the fractional discretization of the \fom, consider the time scale $N_a=\{a,a+1,a+2,...\}$. For $q>0$ and $q\not \in \mathbb{N}$ the $q$-th Caputo-like discrete fractional difference of a function $u:N_a\rightarrow \mathbb{R}$ is defined as is defined \cite{bibnoua,bibcinci} as

\begin{equation}\label{capa}
\Delta_a^q u(t)=\Delta_a^{-(n-q)}\Delta^n u(t)=\frac{1}{\Gamma(n-q)}\sum_{s=a}^{t-(n-q)}(t-s-1)^{(n-q-1)}\Delta^nu(s),
\end{equation}
for $t\in N_{a+n-q}$ and $n=[q]+1$.

$\Delta^n$ is the $n$-th order forward difference operator,
\[
\Delta^n u(s)=\sum_{k=0}^{n}\binom {n}{k}(-1)^{n-k}u(s+k),
\]
while $\Delta_a^{-q}$ represents the fractional sum of order $q$ of $u$, namely,
\begin{equation*}\label{suma}
\Delta_a^{-q}u(t)=\frac{1}{\Gamma(q)}\sum_{s=a}^{t-q}(t-s-1)^{(q-1)}u(s),~t\in \mathbb{N}_{a+q},
\end{equation*}
with the falling factorial $t^{(q)}$ in the following form:
\[
t^{(q)}=\frac{\Gamma(t+1)}{\Gamma(t-q+1)}.
\]

The fractional operator  $\D_a^{-q}$ maps functions defined on $\mathbb{N}_a$ to functions
on $N_{a+q}$ (for time scales see, e.g., \cite{time}).

For the case considered in this paper, $q\in(0,1)$, when $\D u(s)=u(s+1)-u(s)$, $n=1$, and starting point $a=0$, $q$-th Caputo's fractional derivative, $\D ^q$, becomes
\begin{equation}\label{capa}
\D^q u(t)=\frac{1}{\Gamma(1-q)}\sum_{s=a}^{t-(1-q)}(t-s-1)^{(-q)}\Delta u(s).
\end{equation}
Consider next, the FO autonomous IVP in the sense of Caputo's derivative
\begin{equation}\label{trei}
\D^q u(t)=f(u(t+q-1)),~~ t\in \mathbb{N}_{1-q},~~u(0)=u_0,
\end{equation}
with $f$ a continuous real valued map and $q\in(0,1)$. The numerical solution is
\begin{equation}\label{inte}
u(t)=u_0+\frac{1}{\Gamma(q)}\sum_{s=1-q}^{t-q}(t-s-1)^{(q-1)}f(u(s+q-1)),
\end{equation}
or, in the convenient form for numerical simulation \cite{bib1}
\begin{equation}\label{eq3}
u(n)=u(0)+\frac{1}{\Gamma(q)}\sum_{i=1}^n\frac{\Gamma(n-i+q)}{\Gamma(n-i+1)}f(u(i-1)),~{n\in \mathbb{N}^*}.
\end{equation}

The recursive iteration implies that \eqref{eq3} is equivalent to the IVP \eqref{trei} \cite{bibnoua} and, therefore, the study of the IVP \eqref{trei} can be realized, both analytically and numerically, on \eqref{eq3}.

Consider next, the following FO variant of \eqref{trei}, in the sense of Caputo's fractional derivative \eqref{capa}

\begin{equation}\label{primus}
\Delta^q z(t)=f(z(t+q-1),~~t\in N_{1-q}, ~~z(0)=z_0,
\end{equation}
with $f$ a complex continuous function of variable $z=x+\i y \in \mathbb{C}$ and $z_0\in \mathbb{C}$.
Then, the numerical integral \eqref{eq3} of the IVP \eqref{primus} becomes
\begin{equation}\label{ecuss}
 z(n)=z(0)+\frac{1}{\Gamma(q)}\sum_{i=1}^n\frac{\Gamma(n-i+q)}{\Gamma(n-i+1)}f(z(i-1)),~ {n\in \mathbb{N}^*}.
\end{equation}
If one consider the FO variant of the IVP \eqref{zero}, with scaled $c$ within a parametric complex domain, $z(0)=0$, one obtains the numerical integral defining the \fom map

\begin{equation}\label{ecus}
 FO\mathcal{M}:~~z(n)=\frac{1}{\Gamma(q)}\sum_{i=1}^n\frac{\Gamma(n-i+q)}{\Gamma(n-i+1)}f_c(z(i-1)),~ {n\in \mathbb{N}^*},
\end{equation}

    To facilitate understanding the numerical implementation of \eqref{ecus}, with $\Re(f_c(z))=x^2-y^2+c_x$, $\Im(f_c(z))=2xy+c_y$, one gets the following scalar form

\begin{equation}\label{ecus1}
\large
\begin{array}{l}
x(n)= \frac{1}{\Gamma(q)}\mathlarger{\sum}_{i=1}^n\frac{\Gamma(n-i+q)}{\Gamma(n-i+1)}(x(i-1)^2-y(i-1)^2+c_x),\\\\[1mm]
y(n)= \frac{1}{\Gamma(q)}\mathlarger{\sum}_{i=1}^n\frac{\Gamma(n-i+q)}{\Gamma(n-i+1)}(2x(i-1)y(i-1)+c_y),~ {n\in \mathbb{N}^*},
\end{array}
\end{equation}
where $x(0)=y(0)=0$.

If $z(0)$ is variable within a complex domain and $c$ is fixed, one obtains the integral defining the \foj sets
\begin{equation}\label{ecusx}
  FO\mathcal{K}_c:~~z(n)=z(0)+\frac{1}{\Gamma(q)}\sum_{i=1}^n\frac{\Gamma(n-i+q)}{\Gamma(n-i+1)}f_c(z(i-1)),~ {n\in \mathbb{N}^*}.
\end{equation}
The scalar form of \eqref{ecusx} is
\begin{equation}\label{ecusy}
\large
\begin{array}{l}
 x(n)= x(0)+\frac{1}{\Gamma(q)}\mathlarger{\sum}_{i=1}^n\frac{\Gamma(n-i+q)}{\Gamma(n-i+1)}(x(i-1)^2-y(i-1)^2+c_x),\\\\[.01cm]
y(n)= y(0)+\frac{1}{\Gamma(q)}\mathlarger{\sum}_{i=1}^n\frac{\Gamma(n-i+q)}{\Gamma(n-i+1)}(2x(i-1)y(i-1)+c_y),~ {n\in \mathbb{N}^*},
\end{array}
\end{equation}
Note that for the \foj sets, $z(0)$ (i.e. $x(0)$, and $y(0)$ in \eqref{ecusx} and \eqref{ecusy} are variable and scan a complex domain of the variable $z(0)$ (see variables $xx$ and $yy$ in Algorithm 2, Appendix \ref{ap2}).

\section{Properties of the \fom map}
If $c\in$\fom, the orbits generating by \eqref{ecus} are bounded\footnote{Boundedness property of \fom for $c\in \mathbb{C}$ is not analyzed here (The proof, for $c\in \mathbb{R}$, is presented in Proposition \ref{pr4}).}.
Then
$$
\overline{z(n)}=\frac{1}{\Gamma(q)}\sum_{i=1}^{n}\frac{\Gamma(n-i+q)}{\Gamma(n-i+1)}(\overline{z(i-1)}^2+\overline{c}),
$$
with $|\overline{z(n)}|=|z(n)|$, so if  $c\in$\fom then also $\overline{c}\in$\fom.

\noindent Thus, one finds the known property of the \iom set
\begin{proposition}\label{pr1}
\fom set is symmetric about the real axis.
\end{proposition}
Also, for $c$ purely real, following the same reasoning, because $\overline{c}=c$, like for \ioj sets, one has the following property
\begin{proposition}\label{sime}
For $c\in \mathbb{R}$, the \foj sets are symmetric with respect the real axis.
\end{proposition}

Next, for $c\in \mathbb{R}$ one has
$$
z(n)=\frac{1}{\Gamma(q)}\sum_{i=1}^{n}\frac{\Gamma(n-i+q)}{\Gamma(n-i+1)}(z(i-1)^2+c)\ge \frac{c}{\Gamma(q)}\sum_{i=1}^{n}\frac{\Gamma(n-i+q)}{\Gamma(n-i+1)}
$$
Using
\begin{equation}\label{e6}
 \frac{n^q-1}{q}\le\sum_{i=1}^{n}\frac{\Gamma(n-i+q)}{\Gamma(n-i+1)}\le \frac{n^q+1}{q},~~n=1,2,...
\end{equation}
one obtains for $c>0$
$$
z(n)\ge \frac{c}{\Gamma(q)}\frac{n^q-1}{q}\to+\infty\quad n\to+\infty.
$$
Thus, we get
\begin{proposition}\label{pr2}
 For any $c\in \mathbb{R}$, $c>0$, $c\notin$\fom set.
\end{proposition}
Property \ref{pr2} is verified numerically too as indicated by the light blue line AB where $c\in R$ and $c>0$, and $c\notin$ \fom set in Fig. \ref{fig3} (a) for $q=1$ and also Fig. \ref{fig5} (a) for $q=0.5$.

Furthermore, for $c\in \mathbb{R}$ and $c<0$, one obtains
$$
\begin{gathered}
z(0)=0,\\
z(1)=\frac{1}{\Gamma(q)}\frac{\Gamma(1-1+q)}{\Gamma(1-1+1)}(z(0)^2+c)=c,\\
z(2)=\frac{1}{\Gamma(q)}\left(\frac{\Gamma(2-1+q)}{\Gamma(2-1+1)}(z(0)^2+c)+\frac{\Gamma(2-2+q)}{\Gamma(2-2+1)}(z(1)^2+c)\right)\\
=\frac{1}{\Gamma(q)}\left(\frac{\Gamma(1+q)}{\Gamma(2)}c+\frac{\Gamma(q)}{\Gamma(1)}(c^2+c)\right)=qc+c^2+c=c^2+(1+q)c.
$$
\end{gathered}
$$
Assuming
\begin{equation}\label{e7}
c\le -(1+q),
\end{equation}
one obtains $z(2)\ge0$.

\noindent Suppose
$$
\begin{gathered}
z(k)\ge0,\quad k=2,\cdots,n-1,\\
f_c(z(k))\ge\delta,\quad k=1,\cdots,n-1,
\end{gathered}
$$
for $\delta>0$ specified latter, and an $n\ge2$. Then
\begin{equation}\label{e8}
\delta\le f_c(z(1))=z^2+c.
\end{equation}
Next
$$
\begin{gathered}
z(n)=\frac{1}{\Gamma(q)}\Big(\frac{\Gamma(n-1+q)}{\Gamma(n-1+1)}f_c(z(0))+\sum_{i=2}^{n-1}\frac{\Gamma(n-i+q)}{\Gamma(n-i+1)}f_c(z(i-1))\\
+\frac{\Gamma(n-n+q)}{\Gamma(n-n+1)}f_c(z(n-1)\Big)\ge \frac{1}{\Gamma(q)}\Big(\frac{\Gamma(n-1+q)}{\Gamma(n)}c+\frac{\Gamma(q)}{\Gamma(1)}\delta\Big)
\ge \frac{c}{\Gamma(q)}+\delta.
\end{gathered}
$$
If
\begin{equation}\label{e9}
\delta\ge -\frac{c}{\Gamma(q)},
\end{equation}
one obtains $z(n)\ge \frac{c}{\Gamma(q)}+\delta\ge0$.

\noindent Finally, consider
$$
f_c(z(n))\ge f\left(\frac{c}{\Gamma(q)}+\delta\right)=\left(\frac{c}{\Gamma(q)}+\delta\right)^2+c.
$$
Assuming
\begin{equation}\label{e10}
\left(\frac{c}{\Gamma(q)}+\delta\right)^2+c\ge\delta,
\end{equation}
one obtains $f_c(z(n))\ge\delta$.

\noindent If in \eqref{e9} one takes
\begin{equation}\label{e11}
\delta=-\frac{2c}{\Gamma(q)},
\end{equation}
    inserting \eqref{e11} into \eqref{e8} and \eqref{e10}, one obtains
$$
c^2+c\ge -\frac{2c}{\Gamma(q)}\quad\wedge\quad \left(\frac{c}{\Gamma(q)}-\frac{2c}{\Gamma(q)}\right)^2+c\ge-\frac{2c}{\Gamma(q)},
$$
which is equivalent to
$$
c\le-\frac{\Gamma(q)+2}{\Gamma(q)}\quad\wedge\quad c\le-(\Gamma(q)+2)\Gamma(q).
$$
Using $\Gamma(q)\ge1$, the following relations are obtained
$$
-(\Gamma(q)+2)\Gamma(q)\le-(1+q)\quad\wedge\quad-(\Gamma(q)+2)\Gamma(q)\le-\frac{\Gamma(q)+2}{\Gamma(q)}.
$$
Summarizing, if
\begin{equation}\label{e12}
c\le-(\Gamma(q)+2)\Gamma(q),
\end{equation}
then equalities \eqref{e7}, \eqref{e9} and \eqref{e10} hold with \eqref{e11}, and applying the mathematical induction principle, one obtain
$$
\begin{gathered}
z(n)\ge0,\quad \forall n\ge2,\\
f_c(z(n))\ge-\frac{2c}{\Gamma(q)},\quad \forall n\in \mathbb{N}.
\end{gathered}
$$
For $n\ge2$, these estimates along with \eqref{e6} allow to derive
$$
\begin{gathered}
z(n)=\frac{1}{\Gamma(q)}\Big(\frac{\Gamma(n-1+q)}{\Gamma(n-1+1)}f_c(z(0))+\sum_{i=2}^{n}\frac{\Gamma(n-i+q)}{\Gamma(n-i+1)}f_c(z(i-1))\Big)\\
\ge \frac{1}{\Gamma(q)}\Big(\frac{\Gamma(n-1+q)}{\Gamma(n)}c-\frac{2c}{\Gamma(q)}\sum_{i=2}^{n}\frac{\Gamma(n-i+q)}{\Gamma(n-i+1)}\Big)\\
=\frac{1}{\Gamma(q)}\Big(\frac{\Gamma(n-1+q)}{\Gamma(n)}\left(c+\frac{2c}{\Gamma(q)}\right)-\frac{2c}{\Gamma(q)}\sum_{i=1}^{n}\frac{\Gamma(n-i+q)}{\Gamma(n-i+1)}\Big)\\
\ge \frac{1}{\Gamma(q)}\left(c+\frac{2c}{\Gamma(q)}\right)-\frac{2c}{\Gamma(q)}\frac{n^q-1}{q}\to+\infty,
\end{gathered}
$$
as $n\to+\infty$. The above arguments lead to the next result.
\begin{proposition}\label{pr3}
For any $c\in \mathbb{R}$ and $c<0$, satisfying \eqref{e12}, $c\notin$\fom.
\end{proposition}
From Propositions \ref{pr2} and \ref{pr3} one can deduce the boundedness of the \fom for real $c$.
\begin{proposition}\label{pr4}
\fom$\bigcap \mathbb{R}\subset\Big[(-\Gamma(q)+2)\Gamma(q),0\Big)$, i.e., the real part of \fom set is a bounded subset of $\mathbb{R}$.
\end{proposition}

\begin{rremark}\label{remi}
Considering the coefficients $\frac{1}{\Gamma(q)}\frac{\Gamma(n-i+q)}{\Gamma(n-i+1)}$, $n\in \mathbb{N}^*$, clearly one obtains
$$
\lim_{q\to 1}\frac{1}{\Gamma(q)}\frac{\Gamma(n-i+q)}{\Gamma(n-i+1)}=1.
$$
Thus letting $q\to1$ in \eqref{ecuss}, we get
\begin{equation}\label{remunu}
u(n)=u(0)+\sum_{i=1}^{n}f(u(i-1)),~~n\in \mathbb{N},
\end{equation}
which corresponds to iteration of a map $u\to u+f(u)$. Furthermore,
$$
\frac{1}{\Gamma(q)}\frac{\Gamma(n-i+q)}{\Gamma(n-i+1)}=1.
$$
For $i=n$, while for $i<n$, we have
$$
\lim_{q\to 0}\frac{1}{\Gamma(q)}\frac{\Gamma(n-i+q)}{\Gamma(n-i+1)}=\lim_{q\to 0}\frac{1}{\Gamma(q)}\lim_{q\to 0}\frac{\Gamma(n-i+q)}{\Gamma(n-i+1)}=0\frac{1}{n-i}=0.
$$
Thus letting $q\to 0$ in \eqref{ecuss}, we get
\begin{equation}\label{remdoi}
u(n)=u(0)+f(u(n-1)),~~n\in \mathbb{N},
\end{equation}
which corresponds to iteration of a map $u\to f(u)$ only if $u(0)=0$ and, in general, it gives different iterations.  Of course this holds just for finite/bounded times of iterations. Indeed, if $f(u)=1$ for any $u$, from \eqref{ecuss} one obtains
$$
u(n)=u(0)+\frac{1}{\Gamma(q)}\sum_{i=1}^{n}\frac{\Gamma(n-i+q)}{\Gamma(n-i+1)}\sim \frac{n^q}{q},\quad n\to +\infty.
$$
So the iterations do depend on $q$ on $\mathbb{N}$.
\end{rremark}

If the IVP \eqref{primus} is considered in the sense of \fom map (with scanning $c$ and $z(0)=0$), for $q\uparrow 1$,the numerical integral \eqref{remunu} becomes
\begin{equation}\label{remunu1}
z(n)=\sum_{i=1}^{n}f_c(z(i-1)), n\in \mathbb{N},
\end{equation}
while for $q\downarrow 0$
\begin{equation}\label{remunu2}
z(n)=f_c(z(n-1)),~n\in \mathbb{N}.
\end{equation}

If the IVP \eqref{primus} is considered in the sense of \foj sets (with scanning $z(0)$ and fixed $c$), for $q\uparrow 1$, the numerical integral \eqref{remunu} becomes
\begin{equation}\label{remdoi1}
z(n)=z(0)+\sum_{i=1}^{n}f_c(u(i-1)),~~n\in \mathbb{N},
\end{equation}
while for $q\downarrow 0$
\begin{equation}\label{remdoi2}
z(n)=z(0)+f_c(z(n-1)),~~n\in \mathbb{N}.
\end{equation}

\noindent Propositions \ref{pr3} and \ref{pr4} are verified numerically as well. Thus, for $q=0.5$, from \eqref{e12} one gets $c<-6.6865$, values for which \fom does not exists (see Fig. \ref{fig5} (a), wherefrom one can see that the projection of the \fom on the real axis is included in the segment $(-6.6868,0)$. These properties are verified also for the case $q\uparrow1$, when one gets $c<-3$, and where the \fom set does not exists and the projection of the \fom on the real is contained in the segment $(-3,0)$ (Fig. \ref{fig3} (a)).

\begin{rremark}
Studying the dynamics of the case $q=0$ in general is challenging task. For instance, consider the general relation
\begin{equation}\label{e14}
z(n)=z(0)+az(n-1),~~n\in \mathbb{N}.
\end{equation}
for $0\neq a\in \mathbb{C}$ and $z\in \mathbb{C}$. Then
$$
z(n)=z(0)(1+a+\cdots +a^n).
$$
Thus if $|a|>1$ then $\lim_{n\to\infty}u(n)=\infty$. If $|a|<1$, then $\lim_{n\to\infty}z(n)=\frac{z(0)}{1-a}$. The relation \eqref{e14} has the only equilibrium $0$, so the limit point $\frac{z(0)}{1-a}$ is not an equilibrium of \eqref{e14}. Next, taking $\tilde z(0)=\frac{z(0)}{1-a}$, we get the new limit point $\frac{z(0)}{(1-a)^2}$, and repeating this procedure, we get a sequence of limit points $\{\frac{z(0)}{(1-a)^k}\}_{k=1}^\infty$. If $|1-a|<1$ then this sequence tends to $\infty$. If $|1-a|=1$ then this sequence oscillates on the unit circle $S^1$. If $|1-a|>1$ then this sequence tends to $0$.

One can see that the dynamics of \eqref{e14} for $0\ne |a|<1$ is much more complicated then for the standard linear map $u\to au$.
\end{rremark}

\section{Numerical approach of \fom set and \foj sets}\label{mumu}
\fom set and \foj sets are generated with the escape-time algorithm adapted to complex FO discrete equations and is presented in Appendix \ref{ap2}.

In \eqref{ecus}, numerically, $q$ can take the limit value 1 (see Remark \ref{remi})). Therefore, hereafter, the limit case $q\uparrow 1$ will be  considered $q=1$.

Regarding the other limit case, $q\downarrow 0$, due to the fact $\Gamma(0)$ is not defined and the limit is considered on small intervals, in this paper for numerically reasons, the limit has been considered empirically, as $q=1e-15$, and will be denoted $0_+$.

\begin{rremark}
Numerically, the relations \eqref{ecus} with \eqref{remunu1} or \eqref{remunu2} (for $q\uparrow 1$ and $q\downarrow$, respectively) are similar.
Also, the relations \eqref{ecusx} with \eqref{remdoi1} or \eqref{remdoi2} (for $q\uparrow 1$ and $q\downarrow$, respectively) are similar.
\end{rremark}

For $q\downarrow0$, linked to Remark \ref{remi}, one get numerically the following important identity properties (congruent-like shapes) which, for now, are introduced as conjectures

\begin{conj}{1}\label{p5}

For $q\downarrow0$, \fom=\iom. (Fig. \ref{fig4} (a)).
\end{conj}

\begin{conj}{2}\label{p6}

For $q\downarrow0$, and $c=0$, \foj=\iom. (Fig. \ref{fig4} (i)).
\end{conj}

\foj sets are generated for $c\in$\fom, $c\notin$\fom and, for some cases, for $c$ close to \fom neighborhood.
To verify the connectedness property, some empirical zoomed details are considered, where DEM \cite{sci} is used.

In Figs. \ref{fig2} (a)-(d) four representative cases of \fom sets for $q=0_+$, $q=0.25$, $q=0.5$ and $q=1$, respectively, are comparatively presented.
\begin{itemize}

  \item [1.] Consider the \fom for $q=1$ (Figs. \ref{fig3}). As can be seen, there are several differences compared to the IO case. Thus, contrary to expectations, according to which \fom should be similar with \iom while $q$ approaches $1$ (FO cases are considered, generally, that they are generalizations of IO cases), the \fom set for $q=1$ is far from being similar to the \iom set.
Also, the connectedness property of the \iom seems to be broken. For example, the zoom of the detail D in Fig. \ref{fig3} (a), obtained with DEM method (Fig. \ref{fig3} (b)), shows that the \fom seems to not have the connectedness property.
        On the other hand, again contrary to expectations, for $q=1$, \foj sets look similar to IO Julia sets, but translated with $-0.25$ along the real axis. See for example, the $FOK_0$ and $IOK_{0.25}$ (Fig \ref{fig3} (j) and Fig. \ref{fig1} (j)), or $FOK_{-0.25}$ and $IOK_0$ (Fig. \ref{fig3} (i) and Fig. \ref{fig1} (i)). Also, the central symmetry encountered at \ioj sets is verified, but not with respect the origin, but with respect to a horizontally translated center. Like for the \iom set, points $c$ within \fom generate connected \foj (Figs. \ref{fig3} (d), (f), (h-j)), while points outside \fom generate disconnected fractals (Fig. \ref{fig3} (c)). Due to the inherent numerical errors, connectedness property cannot be precisely numerically stated for points close to \fom frontier (see e.g. Fig. \ref{fig3} (e) or (g)).
\item[2.]

Let the other extreme case, $q=0_+$ (Figs. \ref{fig4}). Again, contrary to expectations, in this case, when $q$ approaches $0$, the \fom is similar to \iom (Conjecture \ref{p5}). Empirically verified connectedness property seems to verify (see detail D and his zoom in Fig. \ref{fig4} (a) and (b)). The \foj sets look like some parts of a $\mathcal{M}$ set (probably parts of the \fom) obtained with a magnifier-like. Another property is the fact that, for all values of $c$, within or outside the \fom set, the \foj sets still look connected sets. Also, except the points $c$ on the real axis, the \foj sets have not central symmetry. However, probably the most important property is the fact that for $c=0$, the \foj set is actually the \iom set (see the red rectangle in Fig. \ref{fig4} (i) and Conjecture \ref{p6}).

\item [3.]
For the intermediate case $q=0.5$, see Figs. \ref{fig5}. Now, excepting the values of $c$ belonging to the real axis, where the \foj sets are symmetric with respect the real axis (see Figs. \ref{fig5} (h), (g), (e), (d) and Proposition \ref{sime}), the central symmetry and also connectedness of the \foj are lost (see the zoom of detail D in Fig. \ref{fig5} (a), presented in Fig. \ref{fig5} (b)). Also, the frontier of the \foj is not connected (see the zoomed detail of the region $D_1$, Figs. \ref{fig5} (e) and (f)).
\end{itemize}

\section*{Conclusion and open problems}
In this paper the FO Mandelbrot map and set and FO Julia sets in the Caputo's sense are introduced. Some properties of the \fom map and \foj are analytically studied and some of them numerically verified. The \fom set and \foj sets are obtained with the escape-time algorithm adapted for FO discrete complex maps, while to verify computationally the connectedness properties for some cases, zoomed details are obtained with DEM adapted to FO discrete complex maps. Similarities, and especially differences, between the IO case and FO case are summarizing next.

\begin{figure}
  \includegraphics[width=1\textwidth]{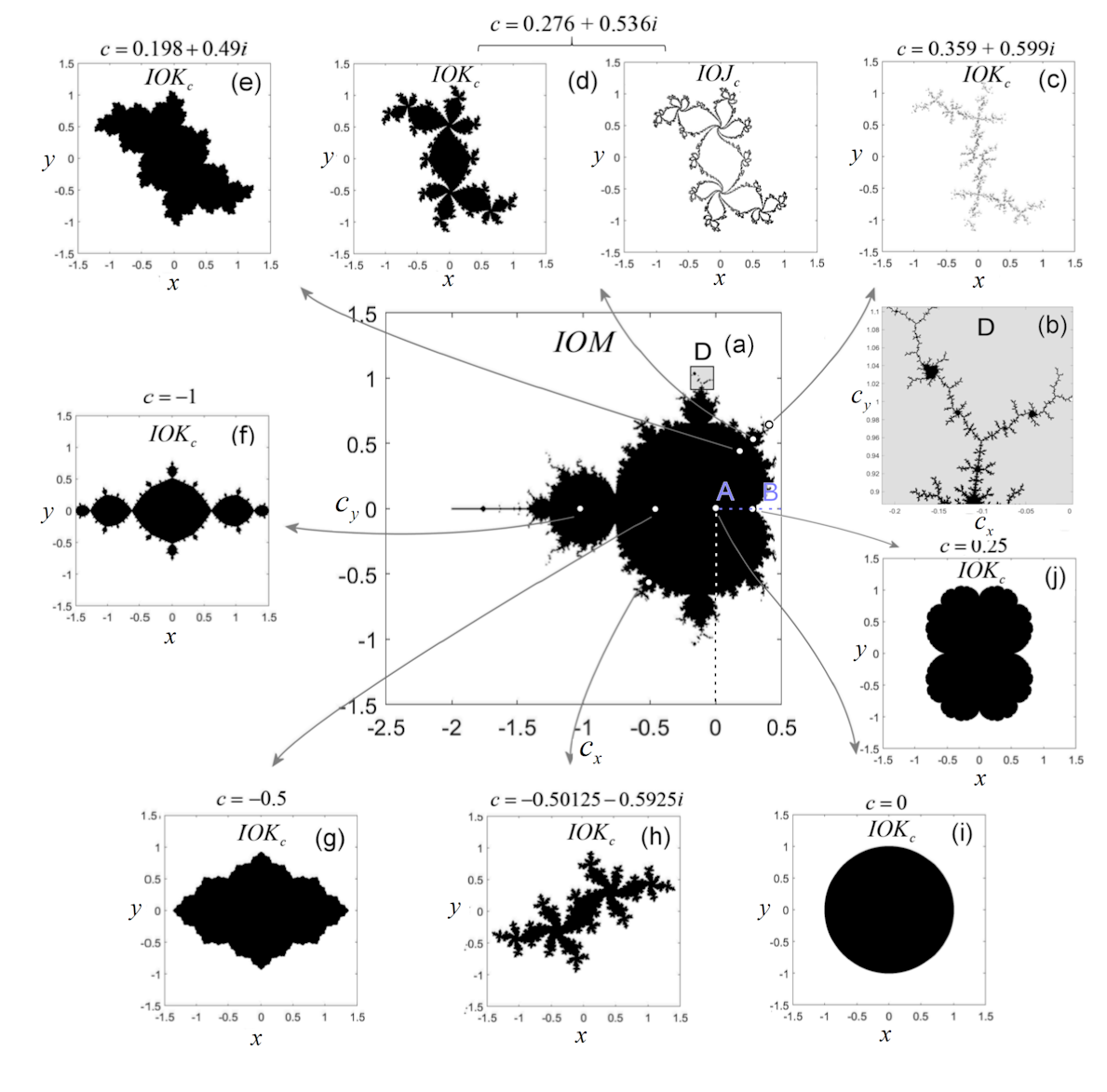}
\caption{The \iom set with few filled Julia sets, \ioj. In Fig. \ref{fig1} (c), the Julia set $IOJ_c$ is presented, while in Fig. \ref{fig1} (b) is presented the zoomed detail D, obtained with the DEM.}
\label{fig1}       
\end{figure}

\begin{figure}
 \includegraphics[width=1\textwidth]{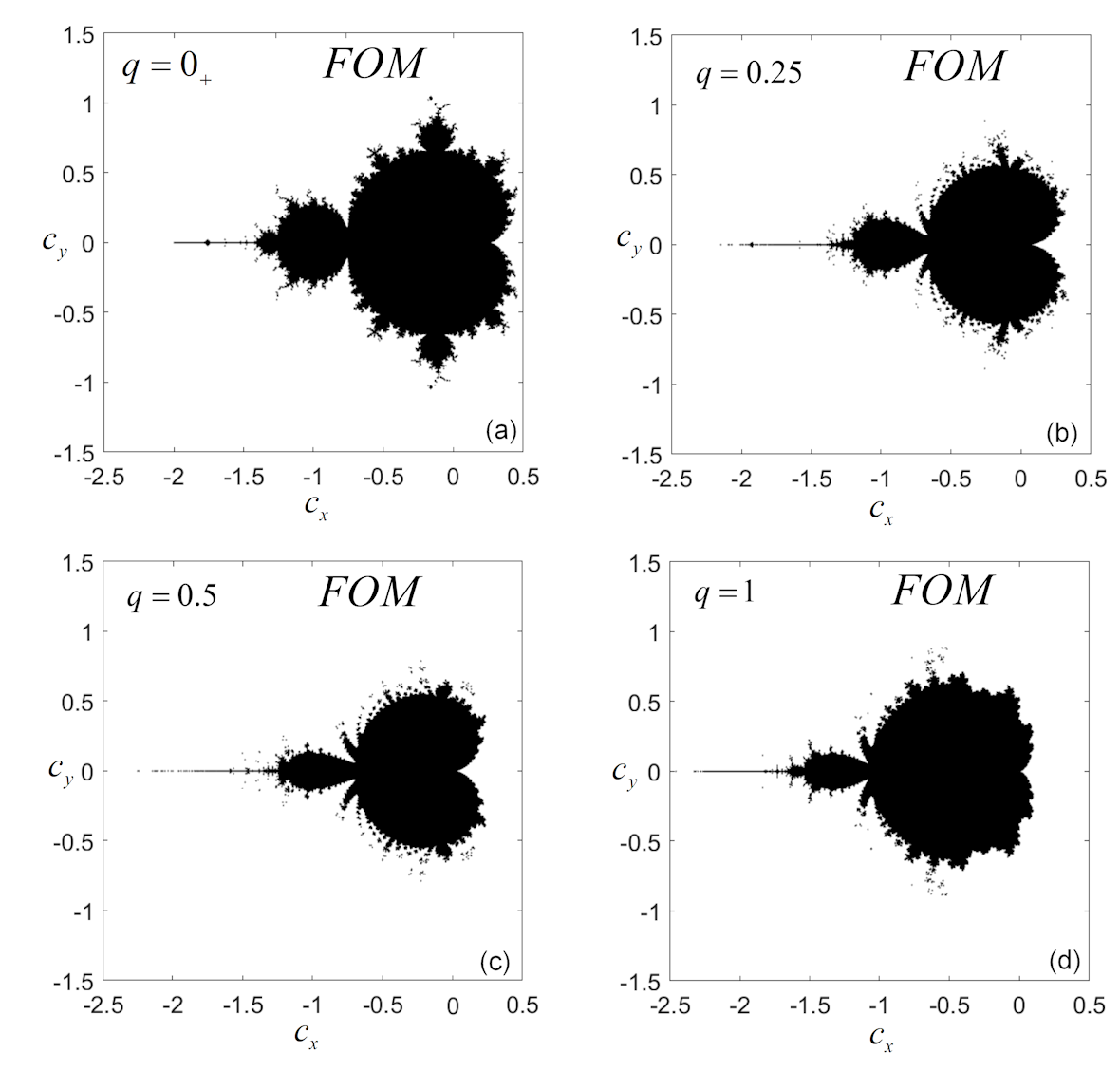}
\caption{Mandelbrot sets: (a) \iom set; (b) \fom set for $q=0.25$; (c) \fom set for $q=0.5$; (d) \fom set for $q=1$.}
\label{fig2}       
\end{figure}

\begin{figure}
  \includegraphics[width=1\textwidth]{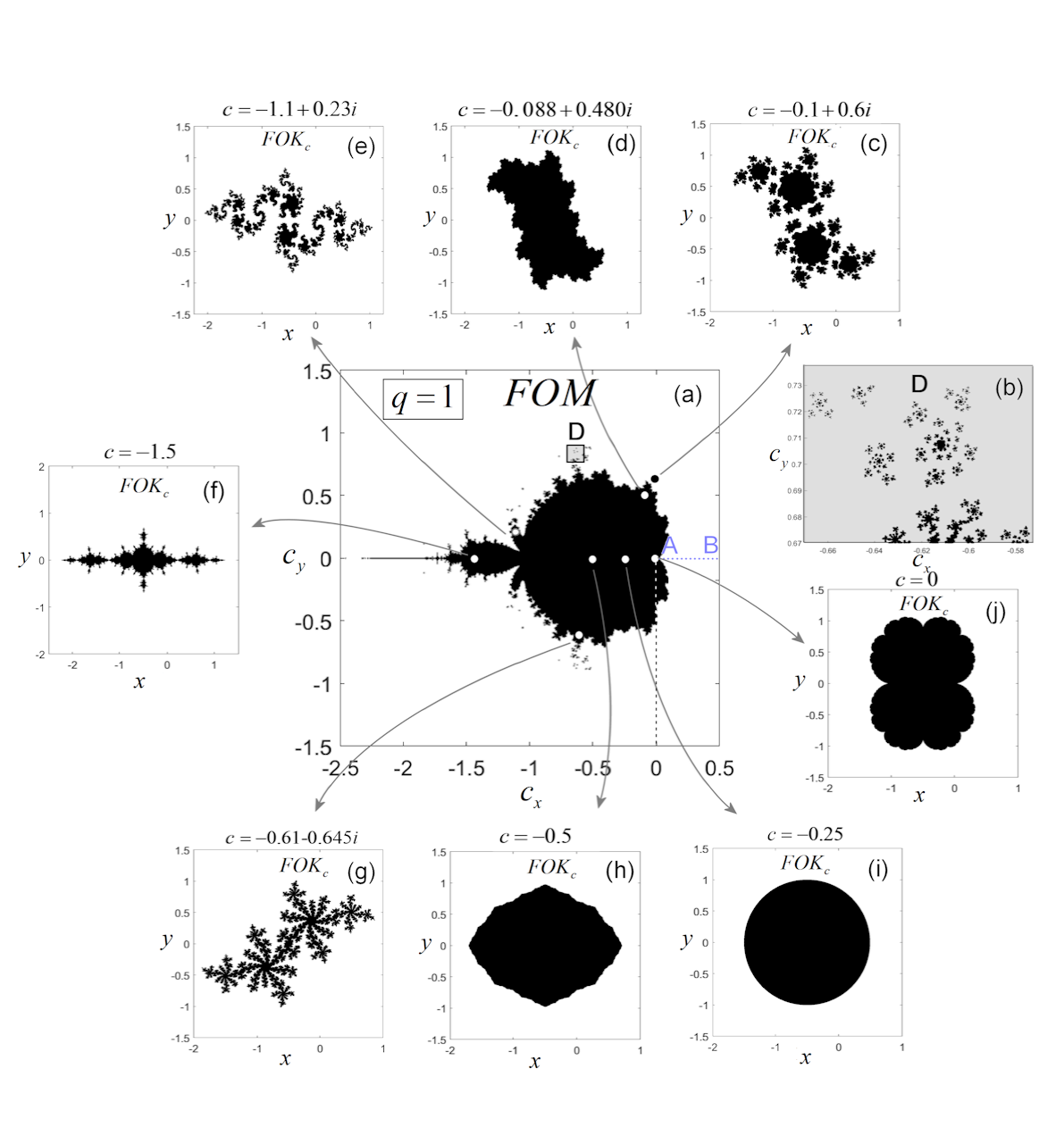}
\caption{\fom set for $q=1$ and some representative \foj sets for different values of $c$; Fig. \ref{fig3} (b) represents the zoomed detail $D$ obtained with DEM indicating disconnectedness of the \fom set for $q=1$.}
\label{fig3}       
\end{figure}

\begin{figure}
  \includegraphics[width=1\textwidth]{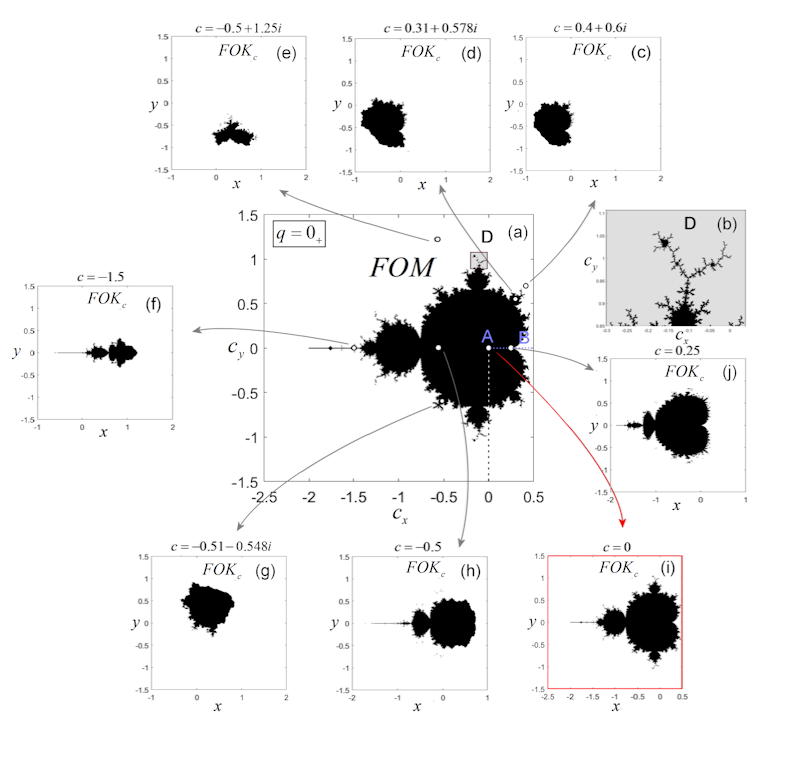}
\caption{\fom set for $q=0_+$ and some representative \foj sets for different values of $c$; Fig. \ref{fig4} (b) represents the zoomed detail $D$ obtained with DEM indicating connectivity property of the \fom for $q=0_+$. For $c=0$ (Fig. \ref{fig4} (i)), the shape of the \foj looks similar to the shape of the \iom set (Fig. \ref{fig1} (a)).}
\label{fig4}       
\end{figure}

\begin{figure}
  \includegraphics[width=1\textwidth]{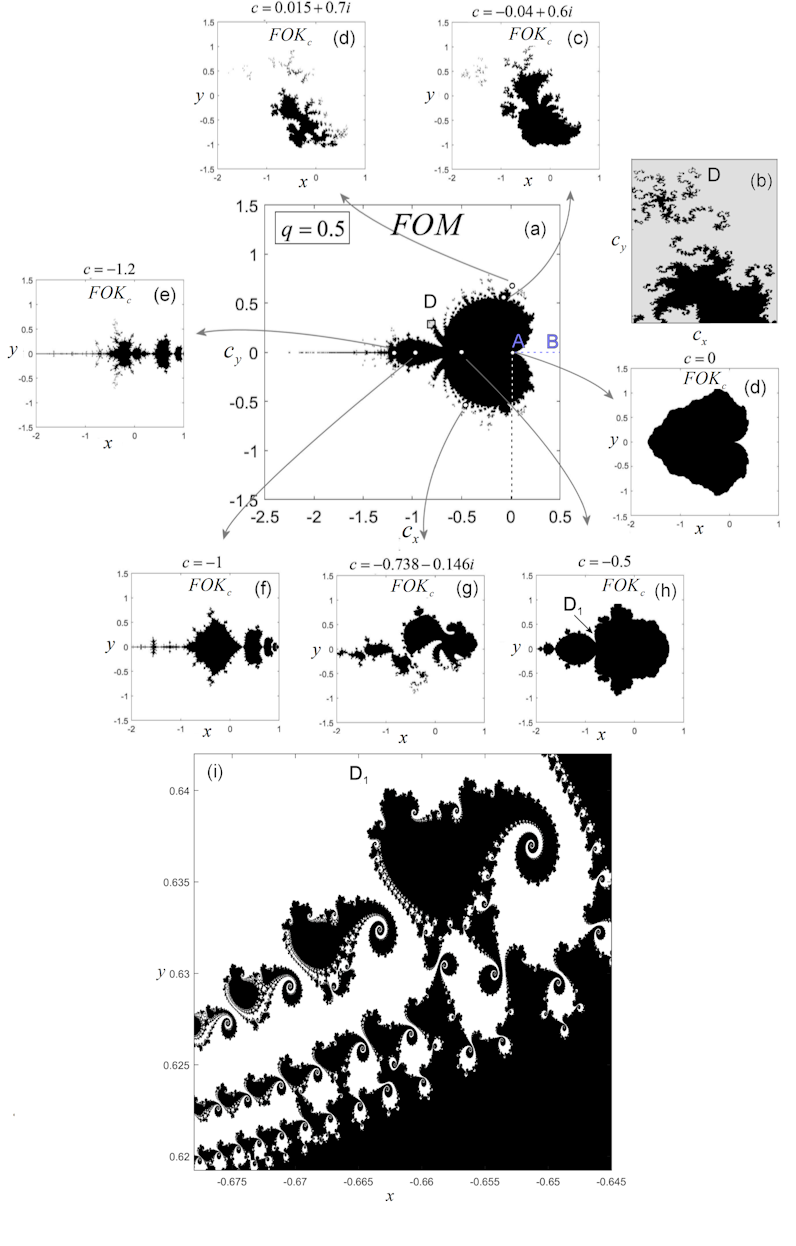}
\caption{\fom set for $q=0.5$ and some representative \foj sets for different values of $c$; Fig. \ref{fig5} (b) represents the zoomed detail $D$ obtained with DEM indicating disconnectedness of the \fom set for $q=0.5$; Fig. \ref{fig5} (f) represents a zoomed detail $D_1$ of the \foj set for $c=-0.5$, indicating the disconnectedness.  }
\label{fig5}       
\end{figure}

\begin{itemize}
\item Probably the most interesting property is the one according to which, for $q\downarrow 0$, the \foj set corresponding to $c=0$, looks similar to the Mandelbrot of IO set (Fig. \ref{fig4}, Conjecture \ref{p6});
\item For all $q$ values, The \fom set and the \foj sets have symmetry with respect the real axis (Proposition \ref{pr1} and Proposition \ref{sime}, respectively);
\item Excepting the case $q=1$, the \foj sets have not central symmetry;
\item For $q=0_{+}$, the underlying \foj corresponding to $c=0$, is not the filled circled, like in the IO case, but the \iom set (see Fig. \ref{fig1} (i) and Fig. \ref{fig4} (i));
\item For $q>0$, the \fom set loose connectedness property (see Paragraphs 1 and Paragraph 3, Section \ref{mumu}), while for $q=0_+$, this property seems verify (see detail D in Figs. \ref{fig4} (a) and (b));
\item For  $q=0_+$ and also for some $q>0$, the \foj sets are imbedded within rectangular domains $\mathcal{L}$ which, contrary to the \ioj sets, are not centered at the origin, but translated along the real axis (see Paragraphs 1 and 3, Section \ref{mumu});
\item The \fom, with $q=1$, is translated horizontally with $-0.25$ (see Figs. \ref{fig1} and Figs. \ref{fig3});
    \item For $q=1$, \foj sets presents a central symmetry with respect a translated center (see Fig. \ref{fig3});
\item Contrary to the IO case, except the case $q=1$, for $q\in(0,1)$, the \foj sets have not central symmetry (see Paragraphs 2 and Paragraph 3, Section \ref{mumu});
\item The FO discretization for the Mandelbrot set, for $q=1$, is not as expected, a generalization of the \iom set (Fig. \ref{fig3}).
\end{itemize}
Beside the results obtained in this paper with the aid of the analytical and numerical approach, several other open problems remain, such as:
\begin{itemize}
 \item the proof of the translation of the \foj sets like in the case $q=1$ (Figs. \ref{fig3});
 \item the fact that for $q\rightarrow 1$ the shape of the \fom set (Fig. \ref{fig3} (a)) does not lead, as expected, to the shape of the \iom set, is typical only to this complex FO system, or apply to other discrete FO systems (real or complex) too? Similar question is open for the case $q\rightarrow 0$;
 \item boundedness of the relations \eqref{ecus}, i.e. boundedness of the \fom set, for the case $c\in \mathbb{C}$ (Proposition \ref{pr4} treats only the case $c\in \mathbb{R}$);
 \item the proof that in the case of $q=1$, the known filled disk for \iom corresponding to the \ioj for $c=0$, looks identical with the filled circled (\foj set) for $c=-0.25$ (Fig. \ref{fig3} (i));
 \item the proof of the identity between the \fom set for $q=0_+$ and the \iom set (Fig. \ref{fig4} (a) and Fig. \ref{fig1} (a), Conjecture \ref{p5});
 \item the proof of the identity between the \foj set for $q=0_+$, $c=0$ and the \iom set (Fig. \ref{fig4} (i), Conjecture \ref{p6});
  \item is the translated center of the symmetry of \foj sets related with the critical point of the \foj sets which, for the \ioj sets of a quadratic polynomial need not be $0$?
\end{itemize}
\vspace{5mm}
\textbf{Acknowledgement}\\
The authors would like to thank Reviewers for taking the necessary time and effort to review the manuscript.
\\
\textbf{Conflict of Interest: }The authors declare that they have no conflict of interest

\clearpage

\clearpage
\section*{Appendices}

\renewcommand{\thesubsection}{\Alph{subsection}}
\renewcommand\thefigure{\Alph{figure}}
\setcounter{figure}{0}
\subsection{Escape-time algoritm for \fom set and \foj sets}\label{ap2}

There are several algorithms to plot the sets as the escape-time method, the boundary scanning method, the inverse iteration method. Also there are several optimizations to increase the speed and images accuracy (see e.g. \cite{sci}). For the exposition clarity, in this paper only black/white coloring scheme is used (for color schemes see e.g. \cite{sci}).

In this paper the Mandelbrot and Julia sets, of IO or FO, are obtained with the slow, but easy to understand escape-time algorithm based on the theorem which states that iterating $f_c$, with starting value $z_0$, only one of the following possibilities happens: either the obtained orbit remains bounded by 2, or diverges to $\infty$. For Mandelbrot set $z_0=0$, and $c$ is varied within a complex parametric domain, usually rectangular, while for Julia sets $z_0$ is varied and $c$ fixed. This well known algorithm is considered in order to facilitate the presentation of the algorithm for the \fom and \foj sets. For simplicity, in this paper the complex parametric plane of $c$ and the plane of the complex variable $z$ are considered similar.

\begin{enumerate}
\item To generate the \fom set with the escape-time algorithm, consider in the cartesian plane the image of a rectangular domain of complex numbers $c$, $\mathcal{L}=\{c_x,c_y| c_x\in[c_{x_{min}},c_{x_{max}}], c_y\in [c_{y_{min}},c_{y_{max}}],c_x,c_y\in \mathbb{R}\}$, with an equidistant grid of $m_x\times m_y$ points $(c_x,c_y)$, $m_x,m_y\in \mathbb{N^*}$. The exploration of numbers $c$ within the considered complex domain $\mathcal{L}$, can be realized with two nested loops, while a third, inner cycle (steps (9)-(13), Fig.\ref{figus1}), the core of the algorithm, makes the escape-time verification. The inside cycle implements the integrals \eqref{ecus1} and \eqref{ecusy}.
As for \iom, for the \fom $\mathcal{L}$ is taken $\mathcal{L}=\{c_x,c_y| c_x\in[-2.5,0.5],c_y\in [-1.5,1.5]\}$.
The domain $\mathcal{L}$ is explored with the steps $step_{c_x}=(c_{x_{max}}-c_{x_{min}})/m_x$, and $step_{c_y}=(c_{y_{max}}-c_{y_{min}})/m_y$.
To generate the \fom, to each $c$ within $\mathcal{L}$, one applies the recurrence \eqref{ecus1} until, either after a chosen finite number of iterations, $N$ (in this paper $N=30$), $|z(n)|$, $n=1,2,...$, $z_0=0$, remains less than 2 and the underlying point $c$ belongs to \fom being plotted black, or $|z(n)|$ becomes greater or equal to 2 (\emph{escape radius}), when $c\not\in$\fom and $c$ is not plotted. Note that because of the symmetry of the \fom, if one intends to generate the entire \fom set, one might save about $50\%$ of drawing time if the algorithm is run only on the superior half of the complex plane, with $y_{min}=0$ and plotting $(c_x,\pm c_y)$.

The pseudocode is presented in Fig. \ref{figus1}.

\item To generate \foj sets, one iterates the map $f_c$, but with $c$ fixed and $z_0$ variable inside $\mathcal{L}$ with the recurrence \eqref{ecusy}. The initial condition, variable, is denoted $z_0:=x+\i y$. If after $N$ iterations, $|z(n)|$, remains less than 2, the underlying point $z_0$ (of coordinates $x$ and $y$) belongs to \foj and is plotted black. If $|z(n)|$ becomes greater or equal to 2, $z_0\not\in$\foj and $z_0$ is not plotted.
The input data are $N$ and data defining $\mathcal{L}$ (${x_{min}},{x_{max}}, {y_{min}}, {y_{max}}$ and $m_x,m_y$), and $c$. For most of \foj sets, $\mathcal{L}=\{x,y| x\in[-2,1],y\in [-1.5,1.5]\}$, not $\mathcal{L}=\{x,y| x\in[-1.5,1.5],y\in [-1.5,1.5]\}$ as for \ioj sets.
The exploration of the domain $\mathcal{L}$ is realized with $step_x=(x_{max}-x_{min})/m_x$, and $step_y=(y_{max}-y_{min})/m_y$.

The pseudocode is presented in Fig. B. variables $xx$ and $yy$ are designed for the inner loop.
\end{enumerate}
Several speed improvements can be done, such as calculating $x(n)\cdot x(n)+y(n)\cdot y(n)$ instead $x^2(n)+y^2(n)$, or calculating only once the expressions $\frac{\Gamma (n-i+q)}{\Gamma(n-i+1)}$ and $\frac{1}{\Gamma(q)}$, or plotting after the domain $\mathcal{L}$ is explored and so on. Also, the algorithm can be written y using e.g. the vectorial calculus, such as the performing matrix calculus of Matlab. Regarding the implementation in Matlab, a solution for the zero index can be found in \cite{x2}.


\label{figus1}
\begin{figure}
\centering
\captionsetup{justification=centering}
\begin{minipage}[t]{.6\textwidth}
\scriptsize
\SetInd{2ex}{1ex}
\begin{algorithm*}[H]
\caption{Algorithm for \fom set}
\SetKwInOut{Input}{Input}
    \Input{$N$\newline$c_{x_{min}},c_{x_{max}},c_{y_{min}},c_{y_{min}}$\newline$m_x$, $m_y$} 
    \vspace{0.1cm}%
$step_{c_x}\gets$ $\frac{c_{x_{max}}-c_{x_{min}}}{m_x}$
\\$step_{c_y}\gets$ $\frac{c_{y_{max}}-c_{y_{min}}}{m_y}$
\\$c_x\gets c_{x_{min}}$
\\  \While {$c_x\leq c_{x_{max}}$}
    {
    $c_y\gets c_{y_{min}}$
        \\\While{$c_y\leq c_{y_{max}}$}
    {$n\gets 1$\\
    \textcolor[rgb]{1.00,0.00,0.00}{$x(0)\gets 0,$
    $y(0)\gets 0$}
     \\\While{$n<N$\bf{and} $x^2(n)+y^2(n)<4$}
     {\vspace{0.1cm}
     $x(n)= \frac{1}{\Gamma(q)}\sum_{i=1}^n\frac{\Gamma(n-i+q)}{\Gamma(n-i+1)}(x^2(i-1)-y^2(i-1)+c_x)$\\ \vspace{0.1cm}
     $y(n)= \frac{1}{\Gamma(q)}\sum_{i=1}^n\frac{\Gamma(n-i+q)}{\Gamma(n-i+1)}(2x(i-1)y(i-1)+c_y)$\\
     \vspace{0.1cm}
     $n\gets n+1$
     \vspace{0.1cm}
     }
     \If{$k=N$}{\bf {plot}$(c_x,c_y,black)$}
             $c_y\gets c_y+step_{c_y}$
    }
    $c_x\gets c_x+step_{c_x}$
    }
\end{algorithm*}
\end{minipage}
\caption{Pseudocode of escape-time algorithm for \fom set. }
\end{figure}

\begin{figure}
\centering
\begin{minipage}[t]{0.6\textwidth}
\scriptsize
\SetInd{2ex}{1ex}
\begin{algorithm}[H]
\caption{Algorithm for \foj set}
\SetKwInOut{Input}{Input}
    \Input{$N$\newline$c_x,c_y$\newline$x_{min},x_{max},y_{min},y_{min}$\newline$m_x$, $m_y$} 
\vspace{0.1cm}%
$step_x\gets$ $\frac{x_{max}-x_{min}}{m_x}$
\\$step_y\gets$ $\frac{y_{max}-y_{min}}{m_y}$
\\$x\gets x_{min}$
\\  \While {$x\leq x_{max}$}
    {
    $y\gets y_{min}$
    \\\While{$y\leq y_{max}$}
    {
        {$n\gets1$\\
        \textcolor[rgb]{1.00,0.00,0.00}{$xx(0)\gets x$, $yy(0)\gets y$}}
        \\\While{$n<N$ \bf{and} $xx^2(n)+yy^2(n)<4$}
  {\vspace{0.1cm}
  $xx(n)= xx(0)+\frac{1}{\Gamma(q)}\sum_{i=1}^n\frac{\Gamma(n-i+q)}{\Gamma(n-i+1)}(xx^2(i-1)-yy^2(i-1)+c_x)$\\\vspace{0.1cm}
     $yy(n)= yy(0)+\frac{1}{\Gamma(q)}\sum_{i=1}^n\frac{\Gamma(n-i+q)}{\Gamma(n-i+1)}(2xx(i-1)yy(i-1)+c_y)$\\
     \vspace{0.1cm}
     $n\gets n+1$
     \vspace{0.1cm}
     }
     \If{$k=N$}{\bf {plot}$(x,y,black)$}
             $y\gets y+step_y$
        }
    $x\gets x+step_x$
    }
\end{algorithm}
\end{minipage}
\caption{Pseudocode of escape-time algorithm for \foj sets. }
\end{figure}



\newpage{\pagestyle{empty}\cleardoublepage}

\end{document}